\documentclass[a4paper,11pt]{article}

\usepackage{contribution}

\newcommand{\weblink}[2][]{%
    \ifthenelse{\equal{#1}{}}%
    {\textnormal{\url{#2}}}%
    {\textnormal{\href{#2}{#1}}}%
}

\newcommand{\acknowledgements}[1]{%
  \bigskip\bigskip
  \textsf{\textbf{\Large Acknowledgements}} \\[2ex]
  {#1}
  \bigskip
}

\def\beq{\begin{equation}}
\def\eeq#1{\label{#1}\end{equation}}
\def\eeqn{\end{equation}}

\def\beqa{\begin{eqnarray}}
\def\eeqa#1{\label{#1}\end{eqnarray}}
\def\eeqan{\end{eqnarray}}

\let\bar=\overbar

\def\Dslash{\not{\hbox{\kern-4pt $D$}}}
\def\dslash{\not{\hbox{\kern-2pt $\del$}}}

\def\msb{{\bar{\ssstyle M \kern -1pt S}}}

\newcommand{\contribution}[7][]{%
  \clearpage
  \thispagestyle{plain}
  \ifthenelse{\equal{#1}{}}
  {\hypersetup{pdftitle={#2}}}
  {\hypersetup{pdftitle={#1}}}
  \hypersetup{pdfauthor={{#3} {#4}}}
  {\centering\normalfont\LARGE\bfseries\sffamily #2 \par\nobreak}
  \lhead{}
  \chead{%
    \textit{\footnotesize XIV International Conference on Hadron Spectroscopy
      (\weblink[\textit{hadron2011}]{http://www.hadron2011.de}), 13-17 June 2011, Munich, Germany}%
  }
  \rhead{}
  \bigskip
  \begin{center}
    {#3} {#4}\ifthenelse{\equal{#6}{}}{}{\footnote{\weblink[#6]{mailto:#6}}}
    \ifthenelse{\equal{#7}{}}{}{#7} \\
    \textit{#5}
  \end{center}
  \bigskip
}

\renewcommand{\abstract}[1]{%
  \begin{center}
    \begin{minipage}{0.85\textwidth}
      \begin{footnotesize}
        #1
      \end{footnotesize}
    \end{minipage}
  \end{center}
  \bigskip
}

\begin{document}

%
%
%
%
%
{  

\makeatletter
\@ifundefined{c@affiliation}%
{\newcounter{affiliation}}{}%
\makeatother
\newcommand{\affiliation}[2][]{\setcounter{affiliation}{#2}%
  \ensuremath{{^{\alph{affiliation}}}\text{#1}}}
%

%

\contribution[$\eta'(958)$ mesonic nuclei]
{$\eta'$ bound states in nuclei and partial restoration of
chiral symmetry}
{Satoru}{Hirenzaki}  
{\affiliation[Department of Physics, Nara Women's University, Nara 630-8506, Japan]{1} \\
 \affiliation[Yukawa Institute for Theoretical Physics, Kyoto University, 
Kyoto 606-8502, Japan]{2} \\
}
{zaki@cc.nara-wu.ac.jp} 
{\!\!$^,\affiliation{1}$, Daisuke Jido\affiliation{2} and Hideko Nagahiro\affiliation{1}}
%
%

\abstract{%
We discuss the in-medium mass of the $\eta^{\prime}$ meson
under partial restoration of chiral symmetry.
The chiral SU(3)$\otimes$SU(3) symmetry tells us 
the flavor singlet pseudoscalar meson $\eta^{\prime}$ should degenerate 
with the octet $\eta$ meson in the SU(3) flavor limit, when chiral symmetry 
is restored in spite of U(1)$_{A}$ anomaly in the flavor single axial current. 
The suppression of the anomaly effect induces an order of 100 MeV 
reduction for the $\eta^{\prime}$ mass at the saturation density without 
introducing a large absorption width. 
We show the formation spectrum of the $\eta'$ mesonic bound state in a nucleus
as a possible observation of the $\eta^{\prime}$ mass reduction.
}
%

\section{Introduction}

Dynamical chiral symmetry breaking and its partial 
restoration in finite density systems is one of the important subjects of 
hadron physics. Recently, spectroscopy of 
deeply bound pionic atom of Sn~\cite{Suzuki:2002ae} and 
low-energy pion-nucleus scattering~\cite{Friedman:2004jh},  
with helps of theoretical analyses~\cite{Kolomeitsev:2002gc}, 
have suggested that the partial restoration does take place in nuclei 
with order of 30\% reduction of the quark condensate. 
The reduction of the quark condensate in nuclear medium also leads to
various phenomena, for instance, 
attractive enhancement of scalar-isoscalar $\pi\pi$ 
correlation in nuclei
and
the suppression of the mass difference between the chiral partners.
Mass reduction of the $\eta^{\prime}$ meson
is also induced by partial restoration of chiral symmetry~\cite{Jido:2011pq}.
The experimental observations of these phenomena, such as
the reduction of the $N$-$N(1535)$
mass difference in the $\eta$ mesonic 
nuclei formation~\cite{Jido:2002yb},
can be further confirmation 
of partial restoration of chiral symmetry in 
nucleus.

\section{$\eta^{\prime}$ mass under chiral symmetry restoration}

Experimentally, a strong mass reduction of $\eta'$ ($\gtrsim 200$ MeV)
has been reported in Ref.~\cite{Csorgo:2009pa} at RHIC.  On the other
hand, a small scattering length ($\sim 0.1$ fm) has been suggested in
Ref.~\cite{Moskal:2000pu} which indicates small mass reduction around 10
MeV at normal saturation density in the linear density approximation.  
The transparency ratio of the $\eta^{\prime}$ meson in nuclei
has suggested the absorption 
width of the $\eta^{\prime}$ 
meson in nuclei is around 30 MeV~\cite{NanovaTalk}. 
Theoretically, NJL model calculations suggested around 200 MeV 
mass reduction
at the saturation 
density~\cite{Costa:2002gk,Nagahiro:2006dr}. In the instanton 
picture, rapid decrease of the effects of instantons in finite energy 
density hadronic matter induces a reduction of the $\eta^{\prime}$
mass~\cite{Kapusta:1995ww}.
An effective model which is consistent to the $\eta' p$ scattering
length data~\cite{Moskal:2000pu} was also proposed
recently~\cite{Oset:2010ub}.  

The basic idea of the present work is that, if density dependence of
the U(1)$_{A}$ anomaly is moderate, a relatively
large mass reduction of the $\eta^{\prime}$ meson is expected 
at nuclear density due to the partial restoration of chiral symmetry~\cite{Jido:2011pq}.
This is based on the following symmetry argument. 
Both the flavor single and octet pseudoscalar mesons composed of 
a $\bar q$-$q$ pair belong to the same 
$(\bf{3},\bf{\bar 3})\oplus (\bf{\bar 3},\bf{3})$ 
chiral multiplet of the SU(3)$_{L}\otimes$SU(3)$_{R}$ group. Therefore,
when the SU(3)$_{L}\otimes$SU(3)$_{R}$ chiral symmetry is manifest,
the flavor singlet and octet mesons should degenerate,
no matter how the U(1)$_{A}$ anomaly effect depends on the density.
In other words, the chiral singlet gluonic current, which makes the 
$\eta^{\prime}$ mass lift up, cannot couple to the chiral pseudoscalar state 
without breaking chiral symmetry.
Hence, the $\eta$ and $\eta^{\prime}$ mass splitting can take place 
only with (dynamical and/or explicit) chiral symmetry breaking, meaning that  
the U(1)$_{A}$ anomaly effect does push the $\eta^{\prime}$ mass up 
but necessarily with the chiral symmetry breaking.
In this way the mass splitting of the $\eta$-$\eta^{\prime}$ mesons is a 
consequence of the interplay of the U(1)$_{A}$ anomaly effect and the 
chiral symmetry breaking. 
Assuming 30\% reduction of the quark condensate in nuclear medium, for instance,
and that the mass difference of $\eta$ and $\eta^{\prime}$ comes 
from the quark condensate linearly, one could expect an order of 100 MeV 
attraction for the $\eta^{\prime}$ meson coming from partial restoration 
of chiral symmetry in nuclear medium. 

The present mechanism of the $\eta^{\prime}$ mass reduction in finite 
density has a unique feature. 
Although some many-body effects introduce an absorptive potential
for the $\eta^{\prime}$ meson in medium,  
the mass reduction mechanism does not involve hadronic intermediate 
states and, thus, the attraction dose not accompany an additional imaginary part. 
Furthermore, in the present case, since the suppression of the U(1)$_{A}$ 
anomaly effect in nuclear medium induces the attractive interaction, 
the influence acts selectively on the $\eta^{\prime}$ meson and, thus, 
it does not induce inelastic transitions of the $\eta^{\prime}$ meson into 
lighter mesons in nuclear medium.  
Consequently 
the $\eta^{\prime}$ meson bound state may have a smaller width
than the binding energy.

\section{Formation spectrum of the $\eta^{\prime}$ mesonic nuclei}

Now we discuss the $\eta^{\prime}$ bound states in a nucleus 
based on the above observation and show expected spectra
of the $\eta^{\prime}$ mesonic nucleus formation in a 
$^{12}$C($\pi^{+},p)^{11}$C$\otimes\eta^{\prime}$ 
reaction~\cite{Jido:2011pq,Nagahiro:2010zz}. 
We perform a simple estimation of the $\eta^{\prime}$  
bound states and, thus, assume a phenomenological optical potential of 
the $\eta^{\prime}$ meson in nuclei as
$
    V_{\eta^{\prime}}(r) = V_{0} \rho(r)/\rho_{0}, 
$    
with the Woods-Saxon type density distribution $\rho(r)$ for nucleus and 
the saturation density $\rho_0=0.17$ fm$^{-3}$. 
The depth of the attractive potential is an order of 100 MeV at the normal nuclear 
density as discussed above and the absorption width is
expected to be less than 40 MeV~\cite{NanovaTalk} which
corresponds to the 20 MeV imaginary part of the optical potential. 
The formation spectrum is calculated in the approach developed  
in Ref.~\cite{Jido:2002yb,Nagahiro:2005gf}
using the impulse approximation and the Green's function method.

\begin{figure}
   \includegraphics[width=0.95\linewidth]{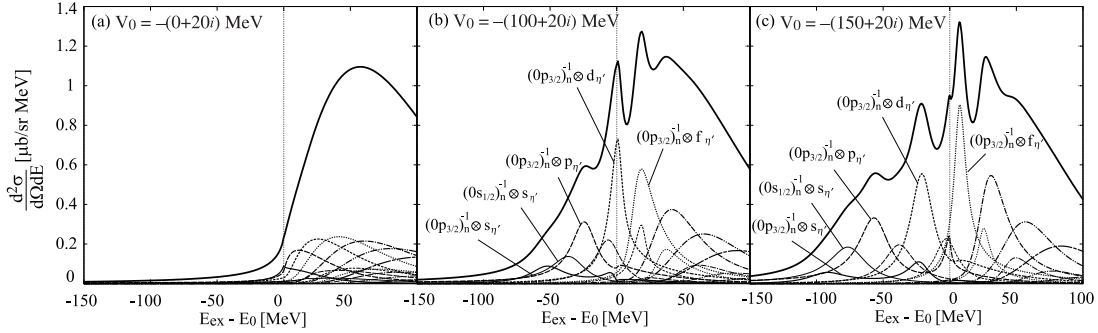}
\caption{{Calculated spectra of the
 $^{12}$C($\pi^+,p)^{11}$C$\otimes\eta'$ at $p_\pi=1.8$ GeV as functions
 of the exitation energy $E_{\rm ex}$ with (a) $V_0=-(0+20i)$ MeV, (b)
 $V_0=-(100+20i)$ MeV and (c) $V_0=-(150+20i)$ MeV.  The thick solid lines
 show the total spectra, and the dominant subcomponents are labeled
 by the neutron-hole state $(n\ell_j)_n^{-1}$ and the $\eta'$ state $\ell_{\eta'}$.
}}
\label{fig:spec}
\end{figure}

In Fig.~\ref{fig:spec}, we show the
calculated $^{12}$C$(\pi^+,p)^{11}$C$\otimes\eta'$ cross
sections with three different potential parameters. 
In the figure, the vertical line 
indicates 
the $\eta^{\prime}$ production threshold in vacuum. 
In the case of no attractive potential, there is no structure in the 
$\eta^{\prime}$-binding  region but some bump in the quasi-free region. 
Finding so prominent peaks in the $\eta^{\prime}$-binding region
as to be possibly observed in future experiments, we conclude that 
with an order of 100 MeV mass reduction and a 40 MeV absorption width 
at the saturation density we have a chance to observe 
the $\eta^{\prime}$-nucleus bound states in the $^{12}$C$(\pi^{+},p)$ reaction.
We see also clear peaks around the $\eta^{\prime}$ production threshold,
for instance $(0p_{3/2})_{n}^{-1}\otimes d_{\eta^{\prime}}$ in plot (b)
and $(0p_{3/2})_{n}^{-1}\otimes f_{\eta^{\prime}}$ in plot (c). They are 
not signals of the bound states, 
however,
these are 
remnants of the bound states which could be formed if the attraction 
would be stronger. Therefore, such peak structure also can be 
signals of the strong attractive potential. 

\section{Conclusion}
We point out that partial restoration of chiral symmetry in a nuclear medium 
induces suppression of the U(1)$_{A}$ anomaly effect to the $\eta^{\prime}$ mass.
Consequently, we expect a large mass reduction of the $\eta^{\prime}$ meson 
in nuclear matter with a relatively smaller absorption width. The mass reduction 
could be observed as $\eta^{\prime}$-nucleus bound states in the formation reactions. 
The interplay between the chiral symmetry restoration 
and the U(1)$_{A}$ anomaly effect can be a clue 
to understand the $\eta^{\prime}$ mass generation mechanism. Therefore,
experimental observations of the deeply $\eta^{\prime}$-nucleus bound states, or 
even confirmation of nonexistence of such deeply bound states,
is important to solve the U(1)$_{A}$ problem.

\acknowledgements{%
This work was partially supported by the Grants-in-Aid for Scientific Research (No. 22740161, No. 20540273, and No. 22105510). This work was done in part under the Yukawa International Program for Quark- hadron Sciences (YIPQS).
}


%

}  


\end{document}